\documentclass[runningheads]{llncs}
\usepackage{subcaption}
\usepackage[T1]{fontenc}
\usepackage{graphicx}
\usepackage{float}
\usepackage{booktabs}
\usepackage{tikz}
\usepackage{pgfplots}
\usepackage{amsmath}
\usepackage{amssymb}
\begin{document}
\title{Machine Learning vs. Randomness: Challenges in Predicting Binary Options Movements}
\titlerunning{Machine Learning vs. Randomness}
%
\author{Gabriel M. Arantes\inst{1}\orcidID{0009-0002-7296-2607}  \and
Richard F. Pinto\inst{1}\orcidID{0009-0007-0176-3383} \and
Bruno L. Dalmazo\inst{1}\orcidID{0000-0002-6996-7602} \and
Eduardo N. Borges\inst{1}\orcidID{0000-0003-1595-7676} \and
Giancarlo Lucca\inst{2}\orcidID{0000-0002-3776-0260} \and
Viviane L. D. de Mattos \inst{1} \orcidID{0000-0002-3512-6290} \and
Fabian C. Cardoso \inst{3} \orcidID{0000-0002-2842-0387} \and
Rafael A. Berri\inst{1}\orcidID{0000-0002-3812-4186}}

\authorrunning{Arantes, G. et al.}
%
\institute{Federal University of Rio Grande (FURG), Rio Grande, Brazil \\
\email{\{gma26062004, richard\_pinto, dalmazo, eduardoborges, vivianemattos,rafaelberri\}@furg.br} \and
Catholic University of Pelotas (UCPel), Pelotas, Brazil \\
\email{giancarlo.lucca@ucpel.edu.br} \and 
University of Rio Verde (UniRV) , Rio Verde, Brazil \\
\email{fabian@unirv.edu.br}}
\maketitle              

\begin{abstract}
Binary options trading is often marketed as a field where predictive models can generate consistent profits. However, the inherent randomness and stochastic nature of binary options make price movements highly unpredictable, posing significant challenges for any forecasting approach. This study demonstrates that machine learning algorithms struggle to outperform a simple baseline in predicting binary options movements. Using a dataset of EUR/USD currency pairs from 2021 to 2023, we tested multiple models, including Random Forest, Logistic Regression, Gradient Boosting, and k-Nearest Neighbors (kNN), both before and after hyperparameter optimization. Furthermore, several neural network architectures, including Multi-Layer Perceptrons (MLP) and a Long Short-Term Memory (LSTM) network, were evaluated under different training conditions. Despite these exhaustive efforts, none of the models surpassed the ZeroR baseline accuracy, highlighting the inherent randomness of binary options. These findings reinforce the notion that binary options lack predictable patterns, making them unsuitable for machine learning-based forecasting.

\end{abstract}
\section{Introduction}\label{sec:introduction}

The financial market has become increasingly complex, requiring advanced tools for analysis and decision-making \cite{twist2020}. In this context, binary options have emerged as financial instruments that offer rapid returns but also carry significant risks due to their speculative nature. Unlike traditional financial assets, where price movements can often be explained by fundamental or technical analysis, binary options operate under conditions that frequently resemble stochastic processes \cite{Silva2020}. This raises the question of whether these instruments can be effectively predicted using data-driven approaches like machine learning.

While ML techniques have shown promise in other financial forecasting tasks \cite{Castilho2021,Obthong2020,Zhang2022}, prior research suggests that markets exhibiting near-random characteristics may fundamentally limit their predictive power \cite{Silva2020}. Given the popularity of binary options and the potential for significant financial losses, it is crucial to investigate whether advanced techniques can provide accurate predictions. This study aims to evaluate the effectiveness of various machine learning techniques in this domain by investigating the impact of feature selection, hyperparameter optimization, and comparing different algorithms (including Random Forest, Logistic Regression, Gradient Boosting, kNN, and neural networks) to assess their ability to learn predictive patterns, using the ZeroR model as a minimum performance benchmark. Our results seek to contribute to the discussion on ML's applicability in highly speculative markets, providing insights into whether these models can forecast binary options or if their limitations reinforce the notion of market randomness.

This paper is structured as follows: Section \ref{sec:Theoretical Foundations} presents the theoretical background. Section \ref{sec:related_works} discusses related works. Section \ref{sec:methodology} describes our methodology, followed by Section \ref{sec:result} with the experimental results. Finally, Section \ref{sec:conclusion} concludes the paper.

\section{Theoretical Foundations}
\label{sec:Theoretical Foundations}

This section briefly covers the core concepts applied in our experiments.

\subsection{Binary Options, Technical Indicators, and ML Algorithms}

Binary options allow traders to forecast an asset's price direction relative to a predefined value at a specific expiration time \cite{Gandar1988}. Their speculative nature and resemblance to stochastic processes pose a significant prediction challenge \cite{Biondo2013,ESMA2018}. We use two widely-acknowledged technical indicators: the Simple Moving Average (SMA) \cite{Brown2005}, a trend-following indicator, and the Relative Strength Index (RSI) \cite{Wilder1978}, a momentum oscillator.

Our study employs several machine learning algorithms: k-Nearest Neighbors (kNN) \cite{1053964}, Random Forest \cite{ho1995random}, Gradient Boosting \cite{FRIEDMAN2002367}, Logistic Regression \cite{10.1001/jama.2016.7653}, Multi-Layer Perceptron (MLP) \cite{jain1996artificial}, and Long Short-Term Memory (LSTM) \cite{hochreiter1997long} networks. The MLP learns complex non-linear dependencies via backpropagation \cite{Rumelhart1986}, with its weight update rule given by:
\begin{equation}
    w_{ij}^{(t+1)} = w_{ij}^{(t)} - \eta \frac{\partial E}{\partial w_{ij}}
\end{equation}
LSTMs are designed to capture long-term temporal patterns, using a memory cell (\(c_t\)) and gates (\(f_t, i_t\)) governed by the following key equations:
\begin{align}
    f_t &= \sigma(W_f \cdot [h_{t-1}, x_t] + b_f) \\
    i_t &= \sigma(W_i \cdot [h_{t-1}, x_t] + b_i) \\
    c_t &= f_t \odot c_{t-1} + i_t \odot \tanh(W_c \cdot [h_{t-1}, x_t] + b_c)
\end{align}

\subsection{Feature Selection, Optimization, and Evaluation}
Feature selection is vital for increasing model efficiency by removing unimportant variables \cite{Guyon2003}. We used the \texttt{SelectFromModel} method with Random Forest to identify the most relevant features based on impurity reduction \cite{Chandrashekar2014}. Hyperparameter optimization refines model performance by tuning learning parameters. We employed Hyperband, an efficient resource-allocation algorithm that adaptively focuses on promising configurations \cite{Li2017}. To ensure reliability, we used k-fold cross-validation for model evaluation, which partitions data into subsets for training and validation to assess generalization \cite{Kohavi1995}. Model performance was measured by accuracy, which quantifies correctly classified instances \cite{Powers2011}.

\section{Related Works}
\label{sec:related_works}

The use of machine learning for financial forecasting has grown in recent years. In deep learning, a notable focus has been on Recurrent Neural Networks for stock price prediction, with studies indicating the potential of models like LSTM and GRU to capture temporal dependencies \cite{chang2024}. Other research has explored network-based features to model market correlation structures, using graph theory to improve prediction accuracy \cite{Castilho2021}.

A significant portion of the literature involves comparing traditional machine learning models with deep learning approaches. For financial time series, deep learning models are often reported to perform better, particularly in capturing long-term patterns \cite{hiransha2018}. Concurrently, comprehensive reviews have highlighted the advantages of fuzzy logic and neural networks in handling market uncertainties \cite{Atsalakis2009}, while early work demonstrated the feasibility of using MLPs, laying the groundwork for their adoption in financial applications \cite{kimoto1990}.

Our research builds on these foundations by systematically applying and evaluating a wide range of these techniques specifically to the binary options market, focusing on the challenges posed by its apparent randomness.

\section{Methodology}\label{sec:methodology}

This section outlines the methodological framework, detailing the dataset, technical indicators, and machine learning techniques. The objective is to test the hypothesis that binary options are entirely random by applying multiple machine learning methods and determining if they can outperform a baseline ZeroR classifier.

\subsection{Dataset Selection and Preprocessing}
We utilized a historical Forex dataset from \textit{HistData}, consisting of minute-by-minute Euro/Dollar (EUR/USD) data from 2021–2023. The dataset was divided into a training set (2021–2022, N = 742,240) and a testing set (2023, N = 322,572) to ensure models learn from historical patterns without information leakage.

\subsection{Feature Engineering}
We derived a set of technical indicators: Simple Moving Averages (SMA) with windows of 2, 3, 4, and 5 minutes (sma2, sma3, sma4, sma5), and Relative Strength Index (RSI) with the same windows (rsi2, rsi3, rsi4, rsi5). These were used as input variables to provide insights into price trends and market momentum.

\subsection{Machine Learning Models and Evaluation}
A variety of machine learning methods were applied: ZeroR (baseline), Random Forest, Logistic Regression, Gradient Boosting, and k-Nearest Neighbors (kNN). A Random Forest-based feature selection process was performed first. To ensure a robust comparison and improve stability, the input features for all models were normalized using the standardization technique (\texttt{StandardScaler}). Each model was evaluated using 5-fold cross-validation on the training data.

\subsection{Hyperparameter Optimization and Neural Networks}
The Hyperband algorithm was used to further investigate model performance by tuning the hyperparameters of the previously tested machine learning models. To further probe for non-linear patterns, several neural network architectures were explored, including Multi-Layer Perceptron (MLP) and Long Short-Term Memory (LSTM) networks. For all neural network experiments, input features were also normalized via \texttt{StandardScaler}. 

A separate experiment was conducted where an MLP was trained on a smaller subset of the original training data to allow for an extended duration (up to 1,000 epochs) with an early stopping mechanism. For this purpose, dedicated training and validation sets were sampled from the 2021--2022 data, while the test set was constructed by combining the remaining 2021--2022 records with the entire original 2023 test set.

\subsection{Final Model Evaluation}
The final evaluation was conducted on the test dataset to assess generalization ability. Accuracy was the main performance metric, enabling direct comparison. The target variable was a binary classification problem: PUT (next price fall) or CALL (next price rise).

\section{Results}\label{sec:result}

This section presents the experimental results, covering feature selection, hyperparameter tuning, and neural network performance.

\subsection{Feature Selection and Initial Performance}
Feature selection using \texttt{SelectFromModel} with a Random Forest classifier retained only the four RSI features (\texttt{rsi2}, \texttt{rsi3}, \texttt{rsi4}, \texttt{rsi5}), discarding all SMA indicators. Table \ref{tab:initial_accuracy} shows that this had a negligible effect on accuracy, suggesting SMA features were not significant. Furthermore, before hyperparameter tuning (Table \ref{tab:initial_hyperparameters}), only Logistic Regression matched the ZeroR baseline (0.5389), while others performed worse.

\begin{table}[H]
    \centering
    \caption{Accuracy before and after feature selection.}
    \begin{tabular}{lcc}
        \toprule
        Model & Before Feature Selection & After Feature Selection \\
        \midrule
        Random Forest & 0.5053 & 0.5062 \\
        \bottomrule
    \end{tabular}
    \label{tab:initial_accuracy}
\end{table}

\begin{table}[H]
    \centering
    \caption{Initial model hyperparameters and accuracy.}
    \begin{tabular}{lcc}
        \toprule
        Model & Initial Hyperparameters & Accuracy \\
        \midrule
        Random Forest & 100 trees & 0.5060 \\
        Logistic Regression & max\_iter=500 & 0.5389 \\
        Gradient Boosting & n\_estimators=100, learning\_rate=0.1 & 0.5386 \\
        kNN & n\_neighbors=5 & 0.5055 \\
        \bottomrule
    \end{tabular}
    \label{tab:initial_hyperparameters}
\end{table}

\subsection{Performance After Hyperparameter Tuning}
After tuning with Hyperband, the optimized hyperparameters (Table \ref{tab:optimized_accuracy}) improved the accuracy of Random Forest, Gradient Boosting, and kNN. However, despite these improvements, no model surpassed the ZeroR baseline of 0.5389, reinforcing the difficulty of predicting binary options.

\begin{table}[H]
    \centering
    \caption{Optimized hyperparameters and accuracy.}
    \begin{tabular}{lcc}
        \toprule
        Model & Optimized Hyperparameters & Accuracy \\
        \midrule
        Random Forest & n\_estimators=161, max\_depth=10, min\_samples\_split=9 & 0.5389 \\
        Logistic Regression & C=0.1699, max\_iter=755, solver=liblinear & 0.5389 \\
        Gradient Boosting & n\_estimators=50, learning\_rate=0.01, max\_depth=3 & 0.5389 \\
        kNN & n\_neighbors=9, metric=manhattan, weights=uniform & 0.5075 \\
        \bottomrule
    \end{tabular}
    \label{tab:optimized_accuracy}
\end{table}

\subsection{Neural Network Performance}

A MLP was tested with the 4 features selected. The architecture consisted of an input layer (4 neurons), two hidden layers (16 and 8 neurons with ReLU), and a sigmoid output layer, with 20\% dropout. Trained for 30 epochs with the Adam optimizer, the MLP achieved an accuracy of 0.5389, identical to the ZeroR model. 

Additionally, an LSTM network was evaluated. The model was configured with an input layer reshaped for sequence data, a single LSTM layer with 16 units, and a Dense sigmoid output layer. Trained for 20 epochs, this model also achieved a final test accuracy of 0.5389.

\subsection{Extended Training Experiment on a Data Subset}

To investigate if an extended training period could enable a model to discover underlying patterns, a final experiment was conducted with a distinct data configuration. A small subset of the 2021-2022 dataset was sampled to create a training set (N = 3,000) and a validation set (N = 3,000). The test set was then constructed by combining the remainder of the 2021-2022 records with the entire original 2023 test set, forming a test set of 1,058,807 samples. The ZeroR baseline accuracy for this specific data partition was calculated to be 0.5379.

For this experiment, a simpler MLP architecture was used, consisting of one hidden layer with 4 neurons and a sigmoid activation function, followed by a single sigmoid output neuron. The model was compiled with the RMSprop optimizer and trained for up to 1,000 epochs, employing an early stopping mechanism with a patience of 200 epochs on the validation loss.

\begin{figure}[H]
    \centering
    \includegraphics[width=0.8\textwidth]{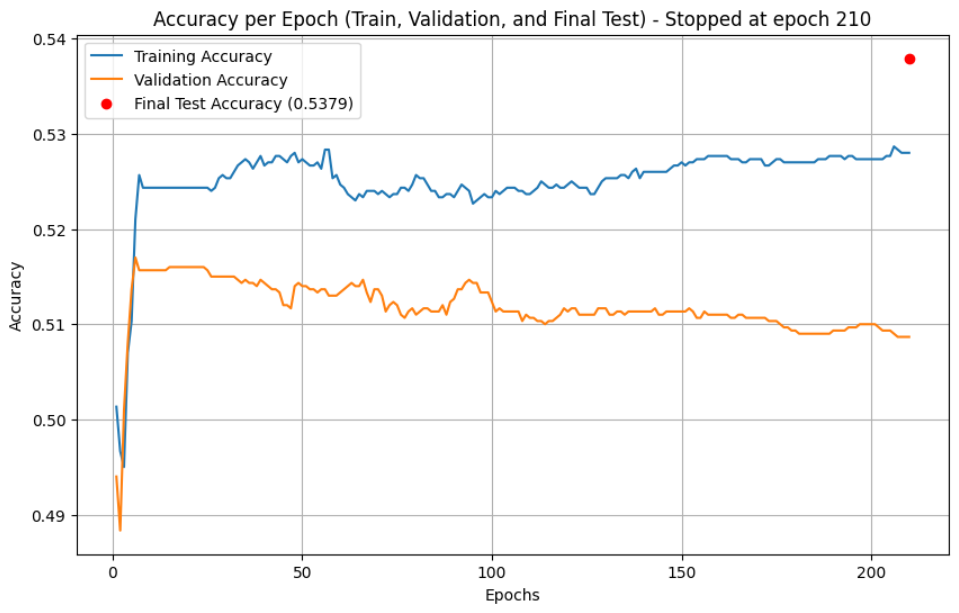}
    \caption{Training and validation accuracy of the MLP over epochs. The divergence between the rising training accuracy (blue) and the decreasing validation accuracy (orange) is a clear sign of overfitting. The final test accuracy on the comprehensive unseen dataset (red dot) matched the ZeroR baseline exactly, confirming that the patterns memorized during training failed to generalize.}
    \label{fig:mlp_extended_train}
\end{figure}

The training was halted by the early stopping callback at epoch 210, with the model restoring the weights from epoch 10, where the best performance on the validation set was observed. The model's performance trajectory, shown in Figure \ref{fig:mlp_extended_train}, reveals a classic case of overfitting. Ultimately, despite the extended training, the MLP's accuracy on the final test set was 0.5379, a result identical to the new ZeroR baseline.

\subsection{Analysis of Model Behavior via Confusion Matrices}

\begin{figure}[H]
    \centering
    \begin{subfigure}{0.3\textwidth}
        \centering
        \includegraphics[width=\linewidth]{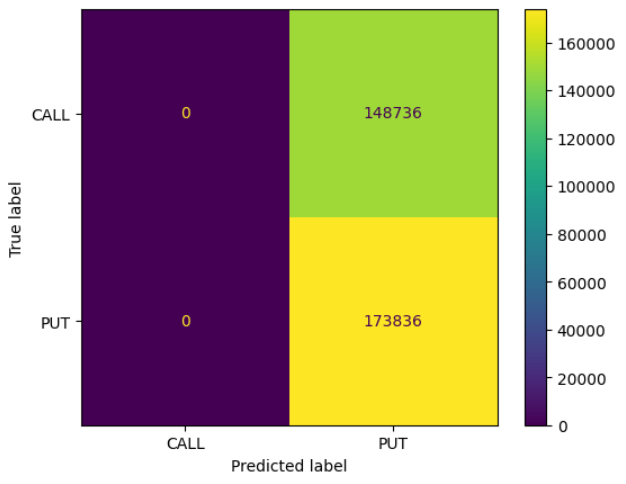}
        \caption{ZeroR (Baseline)}
        \label{fig:cm_zeror}
    \end{subfigure}
    \hfill
    \begin{subfigure}{0.3\textwidth}
        \centering
        \includegraphics[width=\linewidth]{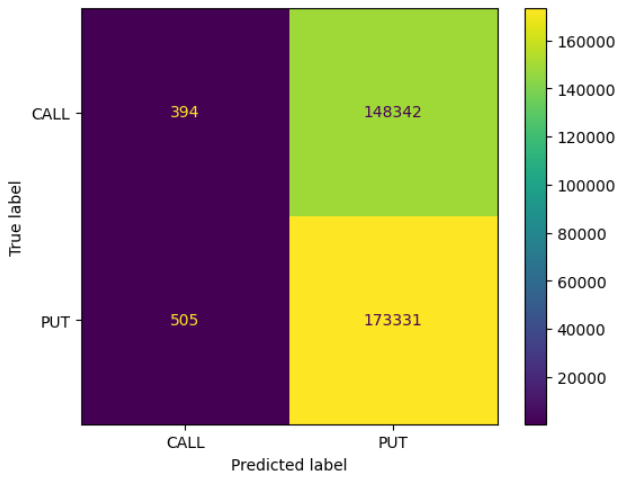}
        \caption{Gradient Boosting}
        \label{fig:cm_gb}
    \end{subfigure}
    \hfill
    \begin{subfigure}{0.3\textwidth}
        \centering
        \includegraphics[width=\linewidth]{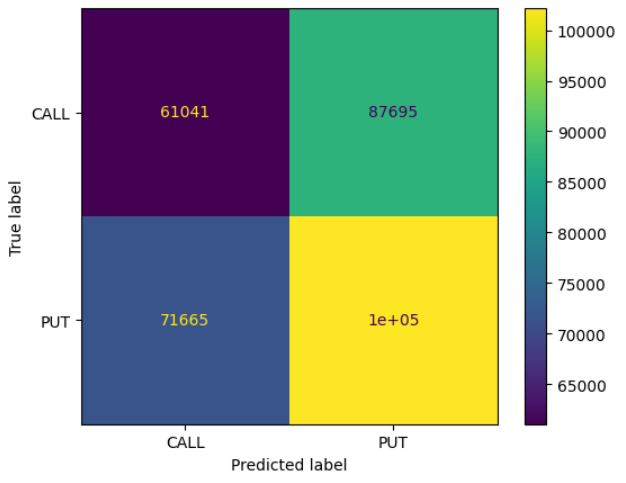}
        \caption{Random Forest}
        \label{fig:cm_rf}
    \end{subfigure}
    \caption{Confusion matrices from models prior to hyperparameter optimization, illustrating three distinct predictive behaviors: (a) the baseline majority-class prediction, (b) a minimal deviation from the baseline, and (c) an unsuccessful attempt to classify both classes.}
    \label{fig:cm_comparison}
\end{figure}

While accuracy scores provide a high-level measure of performance, an analysis of the confusion matrices offers deeper insight into the models' predictive behavior prior to hyperparameter optimization. This analysis reveals why even the more complex models, in their default state, failed to outperform the simple ZeroR baseline. Figure \ref{fig:cm_comparison} illustrates the spectrum of predictive behaviors observed across these initial experiments.

The first and most common pattern was the model's convergence to a strategy indistinguishable from the ZeroR baseline: predicting the majority class ('PUT') exclusively. This was observed in Logistic Regression, the 30-epoch MLP, and the LSTM network (Fig. \ref{fig:cm_comparison}(a)). A slight variation was seen in models like Gradient Boosting and the extended-training MLP, which made a minimal, yet insignificant, attempt to classify the minority class (Fig. \ref{fig:cm_comparison}(b)). The third pattern, from Random Forest and k-NN, involved a more genuine attempt to classify both classes but resulted in high error rates and an accuracy no better than the baseline (Fig. \ref{fig:cm_comparison}(c)).

\subsection{Final Analysis}
\begin{figure}[H]
    \centering
    \includegraphics[width=1\textwidth]{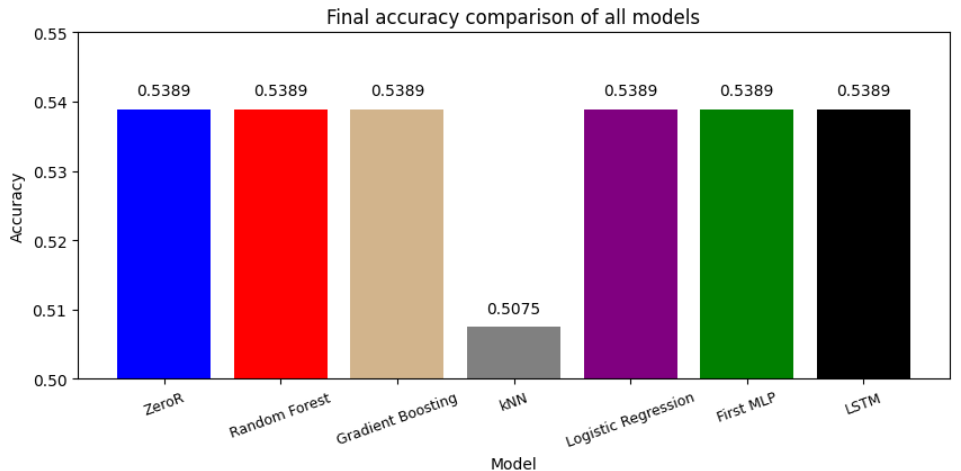}
    \caption{Final accuracy comparison of all primary models against the ZeroR baseline.}
    \label{fig:final_results}
\end{figure}

To summarize all results, Figure \ref{fig:final_results} illustrates the accuracy of each primary model after hyperparameter tuning. Furthermore, the extended training experiment with the MLP, which used a different data partition, also concluded with the model achieving an accuracy identical to its corresponding ZeroR baseline.

As the results from all experiments demonstrate, none of the models outperformed their respective ZeroR baseline. The consistency in accuracy across different algorithms, configurations, and training approaches suggests that binary options exhibit high randomness, making them unsuitable for predictive modeling using these machine learning techniques.

\section{Conclusion}
\label{sec:conclusion}

This study conducted a comprehensive investigation into the applicability of machine learning for predicting binary options movements. A wide range of models was evaluated, from traditional algorithms such as Random Forest, Logistic Regression, Gradient Boosting, and k-NN, to more complex neural network architectures, including Multi-Layer Perceptrons and a Long Short-Term Memory (LSTM) network. The impact of feature selection, extensive hyperparameter optimization, and different training paradigms was systematically examined against a simple ZeroR baseline model.

Our findings are unequivocal: despite the application of diverse and sophisticated learning algorithms, none of the evaluated models were able to outperform their respective ZeroR baseline. A deeper analysis of the models' behavior revealed a critical insight: most algorithms, including the LSTM and MLP, converged to a simplistic strategy of predicting the majority class, as evidenced by their confusion matrices. This demonstrates a fundamental failure to learn any true predictive signal. Even models that attempted a more balanced classification, such as Random Forest, produced high error rates that resulted in even worse performance than the baseline.

The implications of these findings are significant. They reinforce the notion that binary options markets operate with a high degree of stochasticity, posing substantial challenges for machine learning-based forecasting. Our results suggest that simple technical indicators, such as SMA and RSI, are insufficient to capture any underlying predictability. Therefore, future research in this domain should pivot towards fundamentally different approaches. This could involve exploring more holistic feature sets, such as order book information, macroeconomic news, and sentiment data, or employing models specifically designed for noisy, non-stationary environments.

In conclusion, this study demonstrates the profound limitations of applying machine learning techniques to predict binary options movements. The consistent failure of diverse models to find a predictive edge serves as strong empirical evidence supporting the idea of market randomness in this specific, highly speculative context. These results should serve as a cautionary note for both practitioners and researchers, emphasizing that without a source of true predictive information, machine learning models are likely to perform no better than random chance in the specific context of binary options prediction.

\begin{credits}
\subsubsection{\ackname} The authors would like to thank FAPERGS (24/2551-0001396-2, 23/2551-0000773-8), CNPq (305805/2021-5) and FAPERGS/CNPq (23/ 2551-0000126-8).
\end{credits}
%
%
%
\bibliographystyle{splncs04}
\bibliography{bibtex.bib}
\end{document}